\documentclass[showpacs,twocolumn,preprintnumbers]{revtex4}
\usepackage{amssymb}
\usepackage{amsmath}
\usepackage{graphicx}
\usepackage{dcolumn}
\usepackage{bm}

\begin{document}

\title{Full separability criterion for tripartite quantum systems}
\author{Chang-shui Yu}
\author{He-shan Song}
\email{hssong@dlut.edu.cn}
\affiliation{Department of Physics, Dalian University of Technology,\\
Dalian 116024, China}
\date{\today }

\begin{abstract} In this paper, an intuitive approach is employed to
generalize the full separability criterion of tripartite quantum
states of qubits to the higher-dimensional systems (Phys. Rev. A
\textbf{72}, 022333 (2005)). A distinct characteristic of the
present generalization is that less restrictive conditions are
needed to characterize the properties of full separability.
Furthermore, the formulation for pure states can be conveniently
extended to the case of mixed states by utilizing the kronecker
product approximate technique. As applications, we give the analytic
approximation of the criterion for weakly mixed tripartite quantum
states and investigate the full separability of some weakly mixed
states. \end{abstract}
\maketitle

\section{\protect\bigskip Introduction}

Entanglement, as an essential ingredient of quantum information
theory, has been an important physical resource for a lot of quantum
protocols, such as quantum computation [1], quantum cryptography
[2], quantum teleportation [3], quantum dense coding [4] and so on.
Recently, many efforts have been made to characterize the
quantatively properties of entanglement [5-8], however, the good
understanding is only restricted to low-dimensional systems. The
quantification of entanglement for higher dimensional systems and
multipartite quantum systems remains an open question.

Since Coffman et al. [9] introduced the so called residual
entanglement on the basis of concurrence [5], the investigation of
multipartite entanglement has attracted much attention. For example,
D\"{u}r et al. have considered the classification of entanglement
for tripartite systems of qubits [10]; Miyake [11] has given the
classification for multipartite systems based on the
hyperdeterminant. On the basis of the different classes of
multipartite entanglement, the corresponding entanglement monotones
can be given [11,12]. Some quantities have also presented to
characterize the properties of entanglement by collecting the
contributions of the entanglements of different classes [13,14]. One
can note that the quantities introduced in Refs. [13,14] can also
characterize the full separability of a pure multipartite state.
However it is easily found that construction of these quantities
requires more restrictive conditions. Even though some conditions
may be repeated, it is usually not easy to exclude the repeated
ones, especially for high-dimensional systems. Hence, it will reduce
the efficiency of calculation to some extent.

Considering the full separability criterion introduced in Ref. [15],
which can effectively reduce the restrictive conditions to some
extent, in this paper, we will generalize the criterion to
high-dimensional systems by an intuitive approach. The generalized
full separability criterion for pure states can be conveniently
extended to the case of mixed states by utilizing the kronecker
product approximate technique which can usually further reduce
restrictive conditions. As applications, we give the analytic
approximation of the criterion for weakly mixed tripartite quantum
states and study the full separability of some weakly mixed states.
The paper is organized as follows. Firstly, we give the intuitive
generalization of the separability criterion for pure states;
secondly, we extend it to mixed states and discuss the full
separability of some quasi pure states; the conclusions are drawn in
the end.

\section{Full separability criterion for tripartite pure states}

At first, let us recall the full separability criterion for
tripartite pure states of qubits given in Ref. [15]. A tripartite
pure state $\left\vert \psi \right\rangle _{ABC}$ denoted by a
vector in $2\times 2\times 2$
dimensional Hilbert space,%
\begin{equation*}
\left\vert \psi \right\rangle
=(a_{000},a_{001},a_{010},a_{011},a_{100},a_{101},a_{110,}a_{111})^{T},
\end{equation*}%
with the superscript $T$ denoting transpose, is fully separable, if
and only if
\begin{equation}
C(\left\vert \psi \right\rangle )=\left\vert
\boldsymbol{C}(\left\vert \psi \right\rangle )\right\vert
=\sqrt{\sum\limits_{\alpha }\left\vert C^{\alpha }\right\vert
^{2}}=0,
\end{equation}%
here the vector $\boldsymbol{C}(\psi )=\overset{9}{\underset{\alpha =1}{%
\oplus }}C^{\alpha }$ with $C^{\alpha }=$ $\left\langle \psi ^{\ast
}\right\vert s^{\alpha }\left\vert \psi \right\rangle $, where the
star denotes complex conjugation, and
\begin{equation}
s^{1}=-\sigma _{y}\otimes \sigma _{y}\otimes I_{1},s^{2}=-\sigma
_{y}\otimes \sigma _{y}\otimes I_{2},s^{3}=-\sigma _{y}\otimes
I_{1}\otimes \sigma _{y},
\end{equation}%
\begin{equation}
s^{4}=-\sigma _{y}\otimes I_{2}\otimes \sigma
_{y},s^{5}=-I_{1}\otimes \sigma _{y}\otimes \sigma
_{y},s^{6}=-I_{2}\otimes \sigma _{y}\otimes \sigma _{y},
\end{equation}%
\begin{equation}
s^{7}=-\sigma _{x}\otimes \sigma _{y}\otimes \sigma
_{y},s^{8}=-\sigma _{y}\otimes \sigma _{x}\otimes \sigma
_{y},s^{9}=-\sigma _{y}\otimes \sigma _{y}\otimes \sigma _{x},
\end{equation}%
with $\sigma _{x}=\left(
\begin{array}{cc}
0 & 1 \\
1 & 0%
\end{array}%
\right) $, $\sigma _{y}=\left(
\begin{array}{cc}
0 & -i \\
i & 0%
\end{array}%
\right) $, $I_{1}=\left(
\begin{array}{cc}
1 & 0 \\
0 & 0%
\end{array}%
\right) $ and $I_{2}=\left(
\begin{array}{cc}
0 & 0 \\
0 & 1%
\end{array}%
\right) $.

\bigskip As mentioned in Ref. [15], a tripartite pure state of qubits can be
considered as a tensor cubes. Directly, a tripartite
higher-dimensional pure state can naturally considered as a tensor
grid which includes tensor cubes. E.g. let $\left\vert \phi
_{ABC}\right\rangle =\sum_{i,j=0}^{1}\sum_{k=0}^{2}a_{ijk}\left\vert
ijk\right\rangle _{ABC}$, the tensor grid of $\left\vert \phi
_{ABC}\right\rangle $ can be pictured as two adjoining cubes, which
includes three tensor cubes. In this sense, one can draw a
conclusion that tensor cube can be regarded as the unit of tensor
grid. Since every tensor cube in a tensor grid can be considered as
an non-normalized tripartite pure state of qubits, one can get that
every unit corresponds to a $C$ defined in eq. (1). Therefore, the
tensor cube can also be considered as a unit which describes the
full separability of a tripartite higher-dimensional pure state. In
other words, the full separability of the given tripartite
higher-dimensional pure state can be described by the full
separability of the non-normalized tripartite pure state of qubits.

\textbf{Theorem 1:-}For any a tripartite pure state $\left\vert \chi
\right\rangle $ which includes $M$ non-normalized tripartite pure
states of qubits (tensor cubes mentioned above), let the
non-normalized pure state of qubits corresponding to the $i$th cube
be denoted by $\left\vert \varphi _{i}\right\rangle $, one can
obtain the corresponding $C\left( \left\vert \varphi
_{i}\right\rangle \right) $. Define
\begin{equation}
\mathcal{C}(\left\vert \chi \right\rangle
)=\sqrt{\sum_{i=1}^{M}C^{2}\left( \left\vert \varphi
_{i}\right\rangle \right) },
\end{equation}%
for the state $\left\vert \chi \right\rangle $, then $\left\vert
\chi \right\rangle $ is fully separable, if and only if
$F(\left\vert \chi \right\rangle )=0$.

\textbf{Proof.} It is obvious that $\mathcal{C}(\left\vert \chi
\right\rangle )=0$ means that $C\left( \left\vert \varphi
_{i}\right\rangle \right) =0$ holds for all $\varphi _{i}$,
\textit{vice versa}. Since the
tensor cube corresponds to the unit of describing full separability, $%
\mathcal{C}(\left\vert \chi \right\rangle )=0$ shows that there does
not exist any entanglement in $\left\vert \chi \right\rangle $. That
is to say, the tripartite quantum state $\left\vert \chi
\right\rangle $ is fully separable. In other words, since every
non-normalized $\varphi _{i}$ is fully separable, one can obtain
that every group of parallel lines of the tensor grid is linear
dependent. I.e. the state that the grid denotes is fully separable
[15]. On the contrary, if $\left\vert \chi \right\rangle $
is fully separable, $C\left( \left\vert \varphi _{i}\right\rangle \right) =0$%
, i.e. $\mathcal{C}(\left\vert \chi \right\rangle )=0.$

Considering the matrix notation of
\begin{equation*}
\left\vert \chi \right\rangle
=\sum_{i=0}^{n_{1}-1}\sum_{j=0}^{n_{2}-1}\sum_{k=0}^{n_{3}-1}a_{ijk}\left%
\vert ijk\right\rangle ,
\end{equation*}%
$\mathcal{C}(\left\vert \chi \right\rangle )$ can be expressed as
the
function of $\left\vert \chi \right\rangle $, i.e.%
\begin{equation}
\mathcal{C}(\left\vert \chi \right\rangle )=\sqrt{\sum_{\alpha
=1}^{N_{1}}\sum_{\beta =1}^{N_{2}}\sum_{\gamma
=1}^{N_{3}}C^{2}\left( \left( s_{\alpha }\otimes s_{\beta }\otimes
s_{\gamma }\right) \left\vert \chi \right\rangle \right) },
\end{equation}%
where $N_{p}=\frac{n_{p}(n_{p}-1)}{2}$ with $p=1,2,3$; $s_{q}$,
$q=\alpha
,\beta ,\gamma ,$ denotes $2\times n_{p}$ matrix with $p$ corresponding to $%
q $. If the generator of the group $SO(n_{p})$ is denoted by
$S_{p}$, $s_{q}$ can be derived from $\left\vert S_{p}\right\vert $
by deleting the row where all the elements are zero, where
$\left\vert \text{ }\right\vert $ denotes the absolute value of the
matrix elements.

According to eq. (1), eq. (6) can be expanded by

\begin{eqnarray}
\mathcal{C}(\left\vert \chi \right\rangle ) &=&\sqrt{\sum_{\alpha
=1}^{N_{1}}\sum_{\beta =1}^{N_{2}}\sum_{\gamma
=1}^{N_{3}}\sum_{\delta =1}^{9}\left\vert C^{\delta }\left( \left(
s_{\alpha }\otimes s_{\beta }\otimes s_{\gamma }\right) \left\vert
\chi \right\rangle \right)
\right\vert ^{2}}  \notag \\
&=&\left[ \sum_{\alpha =1}^{N_{1}}\sum_{\beta
=1}^{N_{2}}\sum_{\gamma =1}^{N_{3}}\sum_{\delta =1}^{9}(\left\langle
\chi ^{\ast }\right\vert S_{\alpha \beta \gamma }^{T}s^{\delta
}S_{\alpha \beta \gamma }\left\vert
\chi \right\rangle \right.  \notag \\
\times &&\left. \left\langle \chi \right\vert S_{\alpha \beta \gamma
}^{T}s^{\delta }S_{\alpha \beta \gamma }\left\vert \chi ^{\ast
}\right\rangle )\right] ^{1/2},
\end{eqnarray}%
where $S_{\alpha \beta \gamma }=s_{\alpha }\otimes s_{\beta }\otimes
s_{\gamma }$, $s^{\delta }$ are defined by eqs. (2-4), and the superscript $%
T $ denotes transposition operation.

\section{Full separability criterion for mixed states}

On the basis of $\mathcal{C}(\left\vert \chi \right\rangle )$ for
pure
states, the corresponding quantity $\mathcal{C}(\rho )$ for mixed states $%
\rho $ defined in $C_{d\times d}(d=n_{1}\times n_{2}\times n_{3})$
is then given as the convex of
\begin{equation}
\mathcal{C}(\rho )=\inf \sum_{i}p_{i}\mathcal{C}(\left\vert \Psi
_{i}\right\rangle )
\end{equation}%
of all possible decompositions into pure states $\left\vert \Psi
_{i}\right\rangle $ with
\begin{equation}
\rho =\sum_{i}p_{i}\left\vert \Psi _{i}\right\rangle \left\langle
\Psi _{i}\right\vert ,p_{i}\geq 0.
\end{equation}%
$\mathcal{C}(\rho )$ vanishes if and only if $\rho $ is fully
separable.
Substitute eq. (7) into eq. (8), one can get%
\begin{equation}
\mathcal{C}(\rho )=\inf_{U}\sum_{i}p_{i}\left[ \sum_{\alpha
=1}^{N_{1}}\sum_{\beta =1}^{N_{2}}\sum_{\gamma
=1}^{N_{3}}\sum_{\delta =1}^{9}\left\vert C^{\delta }\left(
S_{\alpha \beta \gamma }\left\vert \Psi _{i}\right\rangle \right)
\right\vert ^{2}\right] ^{1/2}.
\end{equation}%
It is obvious that if the infimum of eq. (10) can be provided, one
can obtain a sufficient and necessary condition of separability for
mixed states. However, it seems to be impossible for
higher-dimensional systems. One can only give a lower bound as a
necessary condition. Therefore, a lower bound with strong
sufficiency or convenient for calculations is expected.

According to the matrix notation [7] of equation (9), one can obtain
$\rho
=\Psi W\Psi ^{\dagger }$, where $W$ is a diagonal matrix with $W_{ii}=p_{i}$%
, the columns of the matrix $\Psi $ correspond to the vectors
$\left\vert \Psi _{i}\right\rangle $. Due to the eigenvalue
decomposition: $\rho =\Phi M\Phi ^{\dagger }$, where $M$ is a
diagonal matrix whose diagonal elements are the eigenvalues of $\rho
$, and $\Phi $ is a unitary matrix whose columns are the
eigenvectors of $\rho $, one can obtain $\Psi W^{1/2}=\Phi
M^{1/2}U$, where $U\in C^{r^{\prime }\times N}$ is a Right-unitary
matrix, with $N$ and $r^{\prime }$ being the column number of $\Psi
$ and the rank of $\rho $. Therefore, based on the matrix notation,
eq. (10) can be rewritten as
\begin{eqnarray}
\mathcal{C}(\rho ) &\geqslant &\inf_{U}\sqrt{\sum_{\alpha ,\beta
,\gamma ,\delta }\left\vert U^{T}M^{1/2}\Phi ^{T}S_{\alpha \beta
\gamma }^{T}s^{\delta }S_{\alpha \beta \gamma }\Phi
M^{1/2}U\right\vert _{ii}^{2}}
\notag \\
&=&\inf_{U}\{\sum_{\alpha ,\beta ,\gamma ,\delta }[\left(
U^{T}M^{1/2}\Phi ^{T}S_{\alpha \beta \gamma }^{T}s^{\delta
}S_{\alpha \beta \gamma }\Phi
M^{1/2}U\right)   \notag \\
&&\times \left( U^{\dagger }M^{1/2}\Phi ^{\dag }S_{\alpha \beta
\gamma }^{T}s^{\delta }S_{\alpha \beta \gamma }\Phi ^{\ast
}M^{1/2}U^{\ast }\right) ]_{ii}\}^{1/2},
\end{eqnarray}%
where the Minkowski inequality
\begin{equation*}
\sum\limits_{m}p_{m}\sqrt{\sum\limits_{n}x_{mn}^{2}}\geqslant \sqrt{%
\sum\limits_{n}\left( \sum_{m}p_{m}x_{mn}\right) ^{2}}
\end{equation*}%
is used. According to Ref. [7], one can directly obtain a lower bound of $%
\mathcal{C}(\rho )$ as $\underset{\mathbf{z}}{\text{max}}$ $\tilde{\lambda}%
_{1}(\mathbf{z})-\sum_{i>1}\tilde{\lambda}_{i}(\mathbf{z})$, where $\tilde{%
\lambda}_{j}(\mathbf{z})$ are the singular values of $%
\sum_{j=1}^{I}z_{j}M^{1/2}\Phi ^{T}S_{\alpha \beta \gamma
}^{T}s^{\delta
}S_{\alpha \beta \gamma }\Phi M^{1/2}$ in decreasing order with $\mathbf{z}%
=[z_{1},z_{2},\cdot \cdot \cdot ,z_{I}]$ a group of optimal complex
parameters. It can be easily found that the number of optimal parameters ($%
I=9N_{1}\cdot N_{2}\cdot N_{3}$) is too large to be conveniently
used to calculations for higher-dimensional systems yet. However, it
will be found that by kronecker product approximation technique, not
only might the number of optimal parameters be further reduced, but
also one can calculate the lower bound in different approximation
degrees. In particular, we can provide an analytic approximation for
weakly mixed states.

In fact, if replacing "$\times $" of eq. (11) by "$\otimes $", eq.
(11) can be rewritten as
\begin{eqnarray}
\mathcal{C}(\rho ) &\geq &\inf_{U}\{\sum_{\alpha ,\beta ,\gamma
,\delta }[\left( U^{T}M^{1/2}\Phi ^{T}S_{\alpha \beta \gamma
}^{T}s^{\delta
}S_{\alpha \beta \gamma }\Phi M^{1/2}U\right)   \notag \\
&&\otimes \left( U^{\dagger }M^{1/2}\Phi ^{\dag }S_{\alpha \beta
\gamma }^{T}s^{\delta }S_{\alpha \beta \gamma }\Phi ^{\ast
}M^{1/2}U^{\ast }\right)
]_{ii}^{ii}\}^{1/2}  \notag \\
&=&\inf_{U}\{\left[ \left( U^{T}\otimes U^{\dag }\right) A\left(
U\otimes U^{\ast }\right) \right] _{ii}^{ii}\}^{1/2},
\end{eqnarray}%
where
\begin{equation}
A=\sum_{\alpha ,\beta ,\gamma }\sum_{\delta =1}^{9}\left( \mathbf{\rho }%
^{1/2}\right) ^{T}\mathbf{S}_{\alpha \beta \gamma }^{T}\Sigma ^{\delta }%
\mathbf{S}_{\alpha \beta \gamma }\left( \mathbf{\rho }^{1/2}\right)
,
\end{equation}%
defined in $C_{d\times d}\otimes C_{d\times d}$, and $\mathbf{\rho }%
^{1/2}=\left( \Phi M^{1/2}\right) \otimes \left( \Phi M^{1/2}\right)
^{\ast } $, $\Sigma ^{\delta }=s^{\delta }\otimes s^{\delta },$
$\mathbf{S}_{\alpha \beta \gamma }=S_{\alpha \beta \gamma }\otimes
S_{\alpha \beta \gamma }$. The other indices in above equation are
all defined the same as previous sections. Even though the value of
eq. (11) is not changed, the implied meaning is quite different,
which means that we have copied the given quantum state in a
conjugate Hilbert space and we consider the separability of the
state in a doubled Hilbert space. The distinct advantage is that eq.
(12) allows us to employ the kronecker product approximation
technique [16,17].

Next we will employ the kronecker product approximation technique on
$A$ to derive a lower bound of eq. (12). Based on the technique, $A$
should be converted [19] into $\tilde{A}$ by
\begin{equation*}
\tilde{A}=V_{12}(AV_{12})^{T_{2}},
\end{equation*}%
where the superscript $T_{2}$ denotes partial transposition on the
second subspace [18], $V_{12}$ is swap operator [19] defined as
\begin{eqnarray*}
V_{12} &=&\sum_{ikj^{\prime }k^{\prime }}\delta _{jk^{\prime
}}\delta _{j^{\prime }k}\left\vert j\right\rangle \left\langle
j^{\prime }\right\vert
\otimes \left\vert k\right\rangle \left\langle k^{\prime }\right\vert , \\
j,k^{\prime } &=&1,\cdot \cdot \cdot ,d,j^{\prime },k=1,\cdot \cdot
\cdot ,d.
\end{eqnarray*}%
$\tilde{A}$ has the singular value decomposition:%
\begin{equation}
\tilde{A}=U\Sigma V^{\dag }=\sum_{i=1}^{r}\sigma
_{i}u_{i}v_{i}^{\dag },
\end{equation}%
where $u_{i}$, $v_{i}$ are the $i$th columns of the unitary matrices
$U$ and $V$, respectively; $\Sigma $ is a diagonal matrix with
elements $\sigma _{i}$ decreasing for $i=1,\cdot \cdot \cdot ,r$;
$r$ is the rank of $\tilde{A}$. Thus, based on Ref. [16,17] $A$ can
always be written by

\begin{equation*}
A=\sum_{i}^{r}\mathcal{A}_{i}\otimes \mathcal{B}_{i}=\sum_{i}^{r}\sigma _{i}%
\mathcal{A}_{i}^{\prime }\otimes \mathcal{B}_{i}^{\prime },
\end{equation*}%
where $Vec(\mathcal{A}_{i})=\sqrt{\sigma _{i}}u_{i}$ and $Vec(\mathcal{B}%
_{i})=\sqrt{\sigma _{i}}v_{i}^{\ast }$. For any a $p\times q$ matrix $%
M=[m_{ij}]$ with entries $m_{ij}$ [20], $Vec(M)$ is defined by%
\begin{equation}
Vec(M)=[m_{11},\cdot \cdot \cdot ,m_{p1},m_{12},\cdot \cdot \cdot
,m_{p2},\cdot \cdot \cdot ,m_{1q},\cdot \cdot \cdot ,m_{pq}]^{T}.
\end{equation}%
One can find from eq. (13) that if the two subspace that $A$ is
defined in
is exchanged, $A$ will converted into $A^{\ast }$, hence one has $\mathcal{B}%
_{i}=\mathcal{A}_{i}^{\ast },$ i.e.
\begin{equation}
A=\sum_{i}^{r}\mathcal{A}_{i}\otimes \mathcal{A}_{i}^{\ast
}=\sum_{i}^{r}\sigma _{i}\mathcal{A}_{i}^{\prime }\otimes \mathcal{A}%
_{i}^{\prime \ast }.
\end{equation}
Substitute eq. (16) into eq. (12), eq. (12) can be given by
\begin{equation}
\mathcal{C}(\rho )\geq\inf_{U}\sum_{i}^{N}\sum_{j=1}^{r}\left\vert
\left( U^{T}\mathcal{A}_{j}U\right) _{ii}\right\vert ^{2}.
\end{equation}%
The infimum can be employed to test the full separability of $\rho
$.

In terms of the Cauchy-Schwarz inequality $\left(
\sum\limits_{i}x_{i}^{2}\right) ^{1/2}\times \left(
\sum\limits_{i}y_{i}^{2}\right) ^{1/2}\geqslant
\sum\limits_{i}x_{i}y_{i}$ and $\sum_{i}\left\vert x_{i}\right\vert
\geq \left\vert \sum_{i}x_{i}\right\vert $, $\mathcal{C}(\rho )$
given by eq. (16) can arrive at
\begin{equation}
\mathcal{C}(\rho )\geq \inf_{U}\sum_{i}^{N}\left\vert U^{T}\left(
\sum_{j=1}^{r}z_{j}\mathcal{A}_{j}\right) U\right\vert _{ii},
\end{equation}%
where $z_{j}=x_{j}\exp (i\phi _{j})$, with $x_{j}\geq 0$, $%
\sum_{j}x_{j}^{2}=1.$Therefore the infimum of eq. (18) can be given by $%
\underset{\mathbf{z}}{\text{max}}$ $\lambda _{1}(\mathbf{z}%
)-\sum_{i>1}\lambda _{i}(\mathbf{z})$, where $\lambda
_{j}(\mathbf{z})$ are the singular values of $\left(
\sum_{j=1}^{r}z_{j}\mathcal{A}_{j}\right) $ in decreasing order [7],
with $\mathbf{z}=[z_{1},z_{2},\cdot \cdot \cdot ,z_{r}]$. Note that
$r$ $\leq d^{2}$ is usually much smaller than $d^{2}$ in practical
calculations. In particular, one can consider different numbers of
$\sigma _{j}$ in decreasing order and correspondingly introduce
optimal parameters, which will might provide approximate lower
bounds in different degrees. In this sense, the number of optimal
parameters can be dramatically reduced. In fact, it is very possible
that $\mathcal{A}_{j}$ corresponding to the maximal $\sigma _{j}$
can give the main contribution [8] to the infimum of eq. (18). That
is to say the lower bound of $\mathcal{C}(\rho )$ can be given by
$\lambda _{1}-\sum_{i>1}\lambda _{i}$ with $\lambda _{j}$ the
singular values of $\mathcal{A}_{j}$, which is an analytic
approximation.

For weakly mixed states i.e. quasi pure states, an analytic
approximation of $\mathcal{C}(\rho )$ can also be introduced [21].
According to eq. (13) and kronecker approximation technique, $A$ can
also be given in the following way
\begin{equation*}
A_{l^{\prime }m^{\prime }}^{lm}=\sum_{\alpha ,\beta ,\gamma
}\sum_{\delta =1}^{9}\sqrt{u_{l}u_{l^{\prime }}u_{m}u_{m^{\prime }}}
\end{equation*}%
\begin{equation}
\times \left( \left\langle \Psi _{l}^{\ast }\right\vert S_{\alpha
\beta \gamma }^{T}s^{\delta }S_{\alpha \beta \gamma }\left\vert \Psi
_{l^{\prime }}\right\rangle \times \left\langle \Psi _{m}\right\vert
S_{\alpha \beta \gamma }^{T}s^{\delta }S_{\alpha \beta \gamma
}\left\vert \Psi _{m^{\prime }}^{\ast }\right\rangle \right) ,
\end{equation}%
where $\Psi _{\alpha }$ and $u_{\alpha }$ denote the $\alpha $th
eigenvector and eigenvalue, and all the other quantities are defined
similar to those in eq. (7). According to the symmetry of $A$ given
by eq. (16) and the kronecker product approximation technique in
above section, $A$ can be formally written as
\begin{equation*}
A_{l^{\prime }m^{\prime }}^{lm}=\sum_{\alpha }T_{lm}^{\alpha }\left(
T_{l^{\prime }m^{\prime }}^{\alpha }\right) ^{\ast }.
\end{equation*}%
The density matrix of quasi pure states has one single eigenvalue
$\mu _{1}$ that is much larger than all the others, which induces a
natural order in terms of the small eigenvalues $\mu _{i}$, $i>1$.
Due to the same reasons to
those in Ref. [21], here we consider the second order elements of type $%
A_{11}^{lm}$. Therefore, one can have the approximation
\begin{equation*}
A_{l^{\prime }m^{\prime }}^{lm}\simeq \tau _{lm}\tau _{l^{\prime
}m^{\prime }}^{\ast }\text{ with }\tau
_{lm}=\frac{A_{11}^{lm}}{\sqrt{A_{11}^{11}}}.
\end{equation*}%
In this sense, eq. (18) can be simplified significantly:%
\begin{equation*}
\mathcal{C}(\rho )\simeq \mathcal{C}_{a}(\rho
)=\inf_{U}\sum_{i}\left\vert U^{T}\tau U\right\vert _{ii}.
\end{equation*}%
$\mathcal{C}_{a}(\rho )$ can be given by
\begin{equation*}
\mathcal{C}_{a}(\rho )=\max \{\lambda _{1}-\sum_{i>1}\lambda
_{i},0\},
\end{equation*}%
where $\lambda _{i}$ is the singular value of $\tau $ in decreasing
order.

Consider two $\left( 2\times 2\times 3\right) -$dimensional quasi
pure
states constructed respectively by%
\begin{equation*}
\rho _{1}(x)=x\left\vert GHZ^{\prime }\right\rangle \left\langle
GHZ^{\prime }\right\vert +(1-x)\mathbf{1}_{12}
\end{equation*}%
and
\begin{equation*}
\rho _{2}(x)=x\left\vert W^{\prime }\right\rangle \left\langle
W^{\prime }\right\vert +(1-x)\mathbf{1}_{12},
\end{equation*}%
where
\begin{equation*}
\left\vert GHZ^{\prime }\right\rangle =\frac{1}{2}(\left\vert
000\right\rangle +\left\vert 101\right\rangle +\left\vert
011\right\rangle +\left\vert 112\right\rangle ),
\end{equation*}%
and
\begin{equation*}
\left\vert W^{\prime }\right\rangle =\frac{1}{\sqrt{3}}\left(
\left\vert 000\right\rangle +\left\vert 011\right\rangle +\left\vert
112\right\rangle \right) .
\end{equation*}%
Note that $\left\vert GHZ^{\prime }\right\rangle $ and $\left\vert
W^{\prime }\right\rangle $ given in Ref. [11] correspond to $GHZ$
class and $W$ class with high local rank, respectively. The two
states can be considered as
quasi pure states for $x\geq 0.3$. By the calculation, one can find that $%
\mathcal{C}_{a}(\rho _{1})$ and $\mathcal{C}_{a}(\rho _{2})$ are both \emph{%
nonzero}. What is more, for the quasi pure states generated by the
mixture
of maximally mixed state (identity matrix) and tripartite GHZ state in $%
3\times 3\times 3$ dimension, the corresponding
$\mathcal{C}_{a}(\rho )$s
can all be shown to be \emph{nonzero} for $x\geq 0.3$. We also study some $%
\left( 2\times 2\times 3\right) -$dimensional quasi pure states
$\rho $ by the mixture of maximally mixed state and random
semiseparable pure states generated by Matlab, numerical results
show that $\mathcal{C}_{a}(\rho )$ are $nonzero$ if $\rho $ are
strict quasi pure states. All above show the sufficiency of our
criterion for testing the entanglement of high-dimensional mixed
systems.

\section{Conclusion and Discussion}

In summary, we have utilized an intuitive approach to generalize the
criterion to high-dimensional tripartite systems. The generalized
criterion for pure states can be conveniently extended to the case
of mixed states by utilizing the kronecker product approximate
technique. The lower bound for mixed states can provide necessary
conditions to test the full separability. Compared with the previous
criteria, the criterion introduced here can effectively reduce the
restrictive conditions. However, the criterion is not an
entanglement monotone. Numerical results show that our criterion for
high-dimensional systems is even sufficient condition of full
separability for strict quasi pure states.

\section{Acknowledgement}

This work was supported by the National Natural Science Foundation
of China, under Grant Nos. 10575017 and 60472017.

\end{document}